# Blockchain Based Intelligent Vehicle Data sharing Framework

Madhusudan Singh

Yonsei Institute of Convergence Technology

Yonsei University, Songdo, South Korea

msingh@yonsei.ac.kr

Shiho Kim

School of Integrated Technology

Yonsei University, Seoul, South Korea

shiho@yonsei.ac.kr

## ABSTRACT

The Intelligent vehicle (IV) is experiencing revolutionary growth in research and industry, but it still suffers from many security vulnerabilities. Traditional security methods are incapable to provide secure IV data sharing. The major issues in IV data sharing are trust, data accuracy and reliability of data sharing data in the communication channel. Blockchain technology works for the crypto currency, Bit-coin, which is recently used to build trust and reliability in peer-to-peer networks having similar topologies as IV Data sharing. In this paper, we have proposed Intelligent Vehicle data sharing we are proposing a trust environment based Intelligent Vehicle framework. In proposed framework, we have use the blockchain technology as back bone of the IV data sharing environment. The blockchain technology is provide the trust environment between the vehicles with the based on proof of driving.

## Keywords

Blockchain, intelligent vehicles, security, component; vehicular cloud, ITS

## 1. INTRODUCTION

VANET is the encapsulation of Vehicle-to-Vehicle (V-to-V) and Vehicle-to-Infrastructure (V-to-I), for providing notification of any safety critical incident and hazard to the drivers [1]. This information is gathered by the feedback of the nearby vehicles. This system is prone to security attacks, by marking incorrect feedback, which results in higher congestion and severe hazards [2].

In IV data sharing network, security is a very crucial issue during communication. These networks require trust and privacy [3]. We have proposed a IV-TP element, to build trust and transmit reliable data among IV data sharing. IV-TP is a unique crypto number, which is attached to the message format and transmitted during communication time. The cloud storage based on Blockchain manages the IV-TP, and is accessed ubiquitously. This IV-TP mechanism is also based on Blockchain technology, enabled to create the crypto unique ID, self-executing digital contracts and details of IV, controlled over the Blockchain Cloud [4]. Fig. 1 shows the intelligent vehicle Information sharing environment, showing V2V, V2I communication.

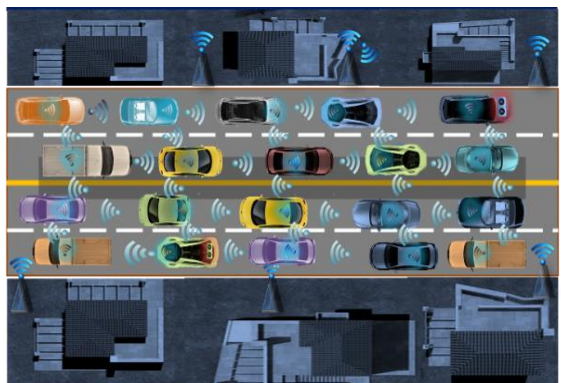

Fig. 1. Intelligent Vehicles Information Environment

Previously, some researchers combined automotive and blockchain technology but most of them considered applications based on services and smart contracts. However, our proposal concentrates on secure and fast communication between intelligent vehicles (self-driving cars) [5]. We have proposed a blockchain based trust environment for intelligent vehicle information sharing based on blockchain technology.

We organize our articles as follows; Section II presents the motivation of using Blockchain based trust environment for data sharing among Intelligent Vehicles over traditional security methods. Section III presents the introduction of blockchain technology and existing work of blockchain technology for Intelligent Vehicles data sharing. Section V, concludes our paper, and discuss our future work for our proposed mechanism.

## 2. MOTIVATION

Current ITS system uses ad-hoc networks for Vehicle communication such as DSRC, WAVE, Cellular Network, which does not guarantee secure data transmission. Currently, vehicle communication application security protocols are based on cellular and IT standard security mechanism which are not up-to-date and suitable for ITS applications. Still many researchers are working to provide standard security mechanism for ITS. Our proposed mechanism is advantages as it is easy to implement, it's a peer -to - peer communication, it provides a secure and trust environment for Vehicle communication with immutable database and ubiquitous data access in a secure way. Our proposal is based on a very simple concept of using Blockchain based trust environment for data sharing among Intelligent Vehicles using the IV-TP (Intelligent Vehicle-Trust Point). We are exploiting the features of Blockchain i.e. distributed and open ledger which is encrypted with Merkel tree and Hash function (SHA-256) and are based on Consensus Mechanism (Proof of Work Algorithm). We have not mentioned the details of the Blockchain mechanism for our application Intelligent Vehicle data sharing due to the limitation of space.

## 3. RELATED WORK

### 3.1. Blockchain Technology

Blockchain technology is distributed, open ledger, saved by each node in the network, which is self-maintained by each node. It provides peer-to-peer network without the interference of the third party. The blockchain integrity is based on strong cryptography that validates and chain blocks together on transactions, making it nearly impossible to tamper with any individual transaction without being detected [6].

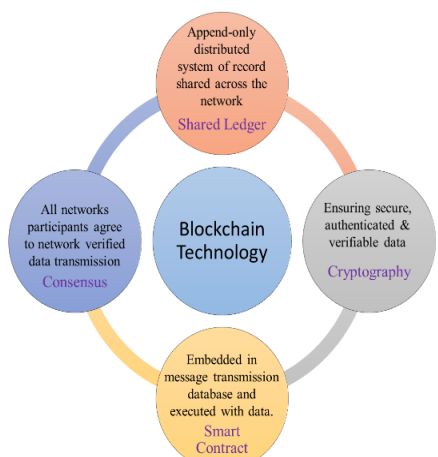

Fig. 2.  Blockchain technology

Fig.2 shows the Blockchain technology features such as shared ledger, Cryptography, Signed blocks of transactions, and digital signatures [6].

### 3.2. Previous work: Blockchain technology for Intelligent Transportation System.

Yong yuan, et.al [7] has proposed the blockchain technology for ITS for establishment of secured, trusted and decentralized autonomous ecosystem and proposed a seven-layer conceptual model for the blockchain.

Benjamin et.al [8], have also proposed the blockchain technology for vehicular ad-hoc network (VANET). They have combined Ethereum' blockchain based smart contracts system with vehicle ad-hoc network. They have proposed combination of two applications, mandatory applications (traffic regulation, vehicle tax, vehicle insurance) and optional applications (applications which provides information and updates on traffic jams and weather forecasts) of vehicles. They have tried to connect the blockchain with VANET services. Blockchain can use multiple other functionalities such as communication between vehicles, provide security, provide peer-to-peer communication without disclosing personal information etc. Ali dorri et.al. [9] have proposed the blockchain technology mechanism without disclosing any private information of vehicles user to provide and update the wireless remote software and other emerging vehicles services. Sean Rowen et.al. [10] have described the blockchain technology for securing intelligent vehicles communication through the visible light and acoustic side channels. They have verified their proposed mechanism through a new session cryptographic key, leveraging both side-channels and blockchain public key infrastructure.

We define our blockchain mechanism for the intelligent vehicles communication environment. We propose a framework for secure trust based environment with peer-to-peer communication between intelligent vehicles without interfering/disturbing other intelligent vehicles.

## 4. BLOCKCHAIN BASED TRUST ENVIORNMENT FOR INTELLIGENT VEHICLES DATA SHARING

We propose a reward based intelligent vehicles communication using blockchain technology. Our proposed mechanism has three basics technologies including communication network enabled connected device, Vehicular Cloud Computing (VCC) and blockchain technology (BT). Fig.3 has shown the complete data-sharing environment for intelligent vehicles.

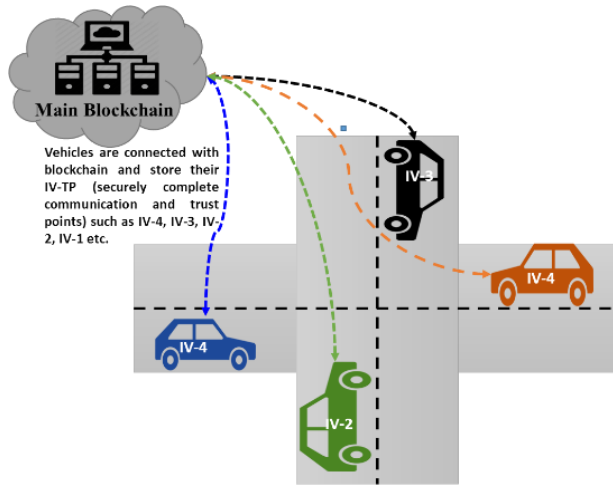

Fig. 3. Proposed blockchain Intelligent Vehicle Communication

## 4.1. Network enabled connected device

It is an internet-enabled device, which can organize, communicate in VANET such as Smartphone, PDA, Intelligent Vehicles, etc.

## 4.2. Vehicular Cloud Computing

VCC is a hybrid technology that has a remarkable impact on traffic management and road safety by instantly using vehicle resources, such as computing, data storage, and internet decision-making.

## 4.3. Blockchain supported intelligent vehicles

Blockchain consists of a technically unlimited number of blocks which are chained together cryptographically in chronological order. In this, each block consists of transactions, which are the actual data to be stored in the chain.

In fig. 6 we present a seven-layer conceptual model for standardizing blockchain architecture for the intelligent communication network. We briefly explain the key features of our proposed network model. Due to space limitation, we did not explain the technical details of the proposed model. The implementation of proposed model is beyond the scope of this paper and thus is omitted.

*Physical layer:* This layer presents the communication network enabled devices such as, IoT devices, mobile, intelligent vehicles, camera, GPS, PDA, etc., which can involve during communication and easily adopt the blockchain mechanism.

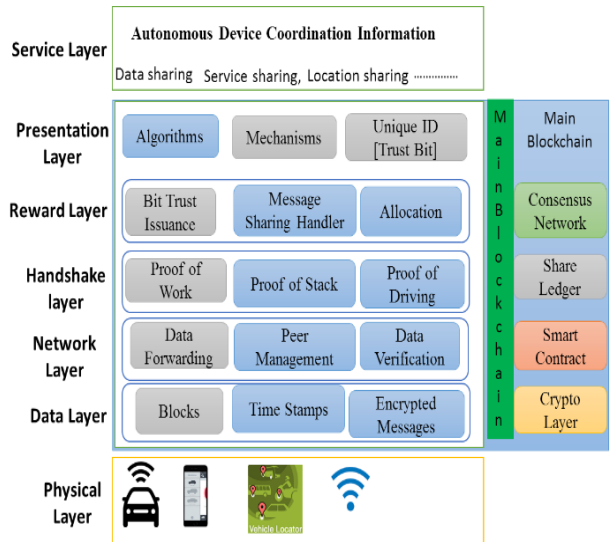

Fig. 4. Proposed Blockchain based Intelligent Vehicle Communication Network Framework

*Data Layer:* This layer process the data blocks with cryptography features such as hash algorithm, merkle tree to make secure blocks.

The structure of block is shown in fig.5, where header part specifies the previous hash and nonce with current hash (root). Hash is made by double SHA 256 algorithm and is not easily hackable.

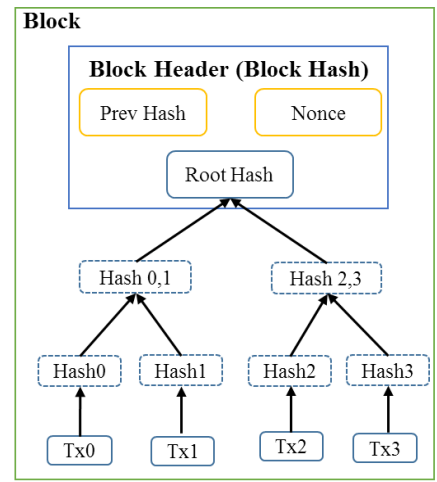

Fig. 5. The structure of blocks.

*Network Layer:* This layer represents the data forwarding peer-to-peer communication and verification of the communication. This layer verifies the legality of the broadcasted message and manages the peer connection between two IVs.

*Handshake Layer:* This feature is called consensus layer in blockchain technology. It provides decentralized communication with in network and helps to develop trust between unknown users in the communication environment. In intelligent vehicles communication networks, more feasible consensus algorithm is proof of driving (PoD), which verify and validate the vehicles involved in the communication networks.

*Reward layer:* It provides some crypto data, which is called IV-TP in this paper. The IV-TP has a crypto data which is assigned to each vehicle and whenever any vehicle wins the consensus competition, its gets some IV-TP from the benefiter IV. The vehicle having the maximum IV-TP, leads in the communication network. IV-TP helps to make trusted environment between the vehicles communication.

*Presentation Layer:* The presentation layer encapsulates multiple scripts, contracts and algorithms, which are provided by the vehicles involved in the network.

*Service Layer:* This layer represents the scenario and use cases of intelligent vehicles communication system.

However, a lot of research organization and startup companies are implementing blockchain in different areas. One such area is, building ITS communication trust environment. Section VI, explains our proposed mechanism through an intersection scenario based use case example.

## 5. CONCLUSION

In this paper, we have presented a reward based intelligent vehicle communication based on blockchain technology and not for specific services as previously proposed by other researchers. We have proposed crypto IV-TP that will help to improve the privacy of IVs. IV-TP provide fast and secure communication between IVs. It also helps to detect the detailed history of IVs communication. IV communication data will be stored on the VC, as long as the user wants. During any accident, the IVs communication history and their reputations are ubiquitously available to authorized organizations (hospital, insurance company, police etc.) and home via VC.

In future, we will simulate our proposed framework mechanism on real-time traffic data of vehicle information sharing scenarios as well as analyze with multiple use cases with a solution.

## 6. ACKNOWLEDGMENT

This work was supported by the MSIP (Ministry of Science, ICT and Future), Korea, under the "ICT Con silience Creative Program" (IITP-2017- 2017-0—1-01 015) supervised by the IITP (Institute of Information & Communications Technology Promotion).